\documentclass[conference,10pt]{IEEEtran}

\usepackage{cite}										
\usepackage{amsmath}									
\usepackage{graphicx}
\usepackage{epstopdf}
\usepackage{float}
\usepackage{bm,upgreek}
\emergencystretch=\maxdimen
\ifCLASSINFOpdf

\else

\fi

\begin{document}
\title{State Estimation in Electric Power Systems Using Belief Propagation: An Extended DC Model}

\author{\IEEEauthorblockN{Mirsad Cosovic}
\IEEEauthorblockA{Schneider Electric DMS NS LLC\\
Novi Sad, Serbia\\
Email: mirsad.cosovic@schneider-electric-dms.com}
\and
\IEEEauthorblockN{Dejan Vukobratovic}
\IEEEauthorblockA{Department of Power, Electronics and Communications Engineering, \\
University of Novi Sad, Serbia\\
Email: dejanv@uns.ac.rs}}

\maketitle

\begin{abstract}
In this paper, we model an extended DC state estimation (SE) in an electric power system as a factor graph (FG) and solve it using belief propagation (BP) algorithm. The DC model comprises bus voltage angles as state variables, while the extended DC model includes bus voltage angles and bus voltage magnitudes as state variables. 
By applying BP to solve the SE problem in the extended DC model, we obtain a Gaussian BP scenario for which we derive closed-form expressions for BP messages exchanged along the FG.
The performance of the BP algorithm is demonstrated for the IEEE 14 bus test case. 
Finally, the application of BP algorithm on the extended DC scenario provides significant insights into a fundamental structure of BP equations in more complex models such as the AC model - the topic we will investigate in our follow up work.
As a side-goal of this paper, we aim at thorough and detailed presentation on applying BP on the SE problem in order to make the powerful BP algorithm more accessible and applicable within the power-engineering community.   
\end{abstract}


\IEEEpeerreviewmaketitle

\section{Introduction}
The state estimation (SE) is an important function of real-time energy management systems (EMS). Typically, SE includes the following processes: network topology processors, observability analysis, state estimation algorithms, and bad data analysis \cite{abur}. The SE algorithm provides an estimate of the state variables according to the network topology and available measurements. The standard approach to obtain the state estimator in electric power systems, formulated as an overdetermined system of non-linear equations, is to solve it as a non-linear weighted least-squares problem \cite{monticelliPaper}.

In view of recent trends in smart grid evolution, there is a growing need for redefining mature algorithms of SE, as well as many other algorithms of EMS, towards distributed and computationally more efficient implementations. In a new, distributed and more dynamic power grid supporting increased number of distributed power sources and time-varying loads, tools emerging in distributed probabilistic systems analysis could provide effective SE solutions. 
    
Probabilistic graphical models seem to be a very good candidate for a more realistic description of an electric power system. In particular, the factor graphs (FGs) possess a potential to bypass many problems of conventional SE. The algorithm for exact inference on probabilistic graphical models without loops is known as the belief propagation (BP) algorithm \cite{pearl}, \cite{bishop}. Using BP algorithm, it is possible to efficiently calculate marginal distributions or a mode of the joint distribution of the system of random variables. The BP algorithm can be also applied to graphical models with loops (loopy BP), although in that case, the solution is not guaranteed to converge to correct marginals/modes of the joint distribution.

To the best of our knowledge, there are only a couple of papers that treat SE through probabilistic graphical models in an electric power system. The work in \cite{kavcic} provides the first demonstration of BP applied to the SE problem. Compared to this work that treats the simplest DC model (thus ignoring reactive power flows and currents), we set our work on more involved extended DC model as an intermediate but insightful step towards deriving BP solution for the AC SE model. The latter is recently addressed in \cite{ilic}, where tree-reweighted BP is applied using preprocessed weights obtained by randomly sampling the space of spanning trees.  

In this paper, we consider an extended DC SE model that we cast into a FG representation and solve using FGs and BP algorithm. The extended DC model is selected not as an alternative for the AC model, but as a simple linear model providing insight into the structure of BP equations. We provide a step-by-step derivation and present a generic format of BP messages in order to make the powerful BP algorithm more accessible and applicable within the power engineering community (which we find missing in \cite{kavcic}, \cite{ilic}). We present numerical results that demonstrate the (loopy) BP performance in simulated IEEE 14 and 30 test case models and comment on the BP convergence. 
\section{Electric power system state estimation}
The SE problem reduces to solving the system of equations:
		\begin{equation}
        \begin{aligned}
        \mathbf{z}=\mathbf{h}(\mathbf{x})+\mathbf{u},
        \end{aligned}
		\label{eqn1}
		\end{equation}
where $\mathbf {x}=(x_1,\dots,x_n)$ is the vector of the state variables, $\mathbf{z} = (z_1,\dots,z_k)$ is the vector of independent measurements (where $n \leq k$), and $\mathbf{u} = (u_1,\dots,u_k)$ is the vector of measurement errors.
		
The nature of measurement functions $\mathbf{h}(\mathbf{x})$ defines the type of the SE: linear functions imply DC SE, while the presence of both non-linear and linear functions imply AC SE. In the electric power system, state variables are bus voltage magnitudes and bus voltage angles, transformer magnitudes of turns ratio and transformer angles of turns ratio. Without loss of generality, in the rest of the paper, we observe bus voltage magnitudes $\mathbf V=(V_1,\dots,V_n)$ and bus voltage angles $\bm \uptheta=(\theta_1,\dots,\theta_n)$ as state variables $\mathbf x$. The measurement errors $\mathbf u$ are assumed to have a zero-mean Gaussian distribution.

The functions $\mathbf{h}(\mathbf{x}) \equiv \mathbf{h}(\mathbf V, \bm \uptheta)$ that connect measurements $\mathbf z$ to state variables $\mathbf x$ are described below.

\textbf{Active and reactive power flow} at the branch that connects buses $i$ and $j$: 
		\begin{equation}
        \begin{aligned}
        h_{P_{ij}}(\cdot) &= g_{ij}V_i^2-V_iV_j(g_{ij}\cos\theta_{ij}+b_{ij}\sin\theta_{ij})\\
        h_{Q_{ij}}(\cdot)  &= -(b_{ij}+b_{si})V_i^2-V_iV_j(g_{ij}\sin\theta_{ij}-b_{ij}\cos\theta_{ij}),
        \end{aligned}
		\label{eqn2}
		\end{equation}
where $V_i$ and $V_j$ are bus voltage magnitudes, while $\theta_{ij} = \theta_i - \theta_j$ is the bus voltage angle difference between buses $i$ and $j$. The parameters in above equations include the conductance $g_{ij}$ and susceptance $b_{ij}$ of the branch, as well as its branch shunt element $b_{si}$. 

\textbf{Active and reactive injection power} into the bus $i$:
		\begin{equation}
        \begin{aligned}
        h_{P_{i}}(\cdot)  &= V_i \sum_{j \in \mathcal{H}} V_j(G_{ij}\cos\theta_{ij}+B_{ij}\sin\theta_{ij})\\
        h_{Q_{i}}(\cdot)  &= V_i \sum_{j \in \mathcal{H}} V_j(G_{ij}\sin\theta_{ij}-B_{ij}\cos\theta_{ij}),
        \end{aligned}
		\label{eqn3}
		\end{equation}
where $\mathcal{H}$ is the set of buses incident to the bus $i$, including the bus $i$. The parameters $G_{ij}$ and $B_{ij}$ are conductance and susceptance of the complex bus matrix.

\textbf{Current magnitude} at the branch connecting buses $i$ and $j$: 
		\begin{equation}
        \begin{gathered}
        h_{I_{ij}}(\cdot) = [aV_i^2+bV_j^2-2V_iV_j(c\cos\theta_{ij}-d\sin\theta_{ij})]^{1/2}\\
        a=g_{ij}^2+(b_{ij}+b_{si})^2,\;\;\;\;
        b=g_{ij}^2+b_{ij}^2\\
        c=g_{ij}^2+b_{ij}(b_{ij}+b_{si}),\;\;\;\;
        d=g_{ij}b_{si}.
        \end{gathered}
		\label{eqn4}
		\end{equation}
		
The equations \eqref{eqn2} - \eqref{eqn4} define functional dependencies of the AC SE model. The AC model is usually approximated (linearised) and, depending on the approximation, different DC models are obtained. In this paper, we focus on the extended DC model \cite{koster}. Similarly as the classical DC model, this model adopts $\theta_{i}-\theta_{j} \approx 0$, which implies $\cos\theta_{ij}\approx1$ and $\sin\theta_{ij}\approx\theta_{ij}$. Unlike the classical DC model, conductance of a branch $g_{ij}$ is non-zero, and without loss of generality, bus voltage magnitudes are $V_i=1+\Delta V_i$ and $V_j=1+\Delta V_j$ . Note that, as compared to DC SE, the extended DC model takes into account both reactive power flows and currents.

With these assumptions and neglecting all quadratic terms, the active and reactive power flow equations \eqref{eqn2} reduce to:
		\begin{equation}
        \begin{aligned}
        h_{P_{ij}}(\cdot) &= g_{ij}(V_i-V_j)-b_{ij}\theta_{ij}\\
        h_{Q_{ij}}(\cdot) &= -(b_{ij}+2b_{si})V_i+b_{ij}V_j-g_{ij}\theta_{ij}+b_{si}.
        \end{aligned}
		\label{eqn5}
		\end{equation}
		
The injection active and reactive power \eqref{eqn3}, reduces to:  
		\begin{equation}
        \begin{aligned}
        h_{P_{i}}(\cdot) =& G_{ii}V_i + \sum_{j \in  \mathcal{H} \setminus i}B_{ij}\theta_i+
        \sum_{j \in  \mathcal{H} \setminus i}(G_{ij}V_j-B_{ij}\theta_{j})\\
        h_{Q_{i}}(\cdot) = &-(2B_{ii}+{\sum_{j \in  \mathcal{H} \setminus i}}B_{ij})V_i + \sum_{j \in  \mathcal{H} \setminus i}G_{ij}\theta_i\\ 
        &-\sum_{j \in  \mathcal{H} \setminus i}(B_{ij}V_j+G_{ij}\theta_j)+\sum_{j \in  \mathcal{H}}B_{ij},
        \end{aligned}
		\label{eqn6}
		\end{equation}		
where $ \mathcal{H}\setminus i$ is the set of buses incident to the bus $i$.

The relation for current magnitude \eqref{eqn4} is transformed into a linear equation conditioned that $b_{si}=0$:		
		\begin{equation}
        \begin{gathered}
        h_{I_{ij}}(\cdot) = \sqrt{(g_{ij}^2+b_{ij}^2)}|V_i-V_j|.\\
        \end{gathered}
		\label{eqn7}
		\end{equation}
		
The system of equations \eqref{eqn5} - \eqref{eqn7} defines a set of extended DC model measurement functions. Note that, unlike classical DC model where state variables include only bus voltage angles, the extended DC model takes into account bus voltage magnitudes.

Under the assumption that measurement errors follow zero-mean Gaussian distribution, the probability density function associated with the m-th measurement:
		\begin{equation}
        \begin{gathered}
        \mathcal{N}(z_m|\mathbf{x},\sigma_m^2) = \cfrac{1}{\sigma_m\sqrt{2\pi}} 
        \exp\Bigg\{\cfrac{[z_m-h_m(\mathbf{x})]^2}{2\sigma_m^2}\Bigg\},
        \end{gathered}
		\label{eqn8}
		\end{equation}
where $z_m$ is the value of the measurement, $\sigma_m^2$ is the measurement variance, and the function $h_m(\mathbf{x})$ connects the vector of state variables to the value of the m-th measurement.

One can find the maximum a posteriori probability (MAP) solution to the SE problem via maximization of the likelihood function, which is defined via likelihoods of $k$ independent measurements:  
		\begin{equation}
        \begin{gathered}
		\mathrm{arg} \max_{\mathbf{x}}\mathcal{L}(\mathbf{z}|\mathbf{x}) 
		= \prod_{h=1}^k \mathcal{N}(z_h|\mathbf{x},\sigma_h^2).
        \end{gathered}
		\label{eqn9}
		\end{equation}
The conventional SE is using weighted least-squares to solve the optimization problem defined in \eqref{eqn9}.				

\section{Factor Graphs and Belief propagation algorithm}  
As in many fields, SE in an electric power system deals with the problem of determining state variables $\mathbf{x}$ according to the noisy observed data $\mathbf{z}$ and some prior knowledge: 
		\begin{equation}
        \begin{gathered}
		p(\mathbf{x}|\mathbf{z})=\cfrac{p(\mathbf{z}|\mathbf{x})p(\mathbf{x})}{p(\mathbf{z})}.
        \end{gathered}
		\label{eqn10}
		\end{equation}
Assuming that the prior probability distribution $p(\mathbf{x})$ is uniform, and given that $p(\mathbf{z})$ is a constant, the most probable or MAP solution of \eqref{eqn10} reduces to the maximum likelihood solution, as given below \cite{barber}:
		\begin{equation}
        \begin{gathered}
		\hat{\mathbf{x}}=\mathrm{arg}\max_{\mathbf{x}}p(\mathbf{x}|\mathbf{z})=\mathrm{arg}\max_{\mathbf{x}}p(\mathbf{z}|\mathbf{x})=\mathrm{arg}\max_{\mathbf{x}}\mathcal{L}(\mathbf{z}|\mathbf{x}).
        \end{gathered}
		\label{eqn11}
		\end{equation}
If $\mathcal{L}(\mathbf{z}|\mathbf{x})$ can be factorized into factors affecting small subsets of state variables $\mathbf{x}$, which is the case as given in \eqref{eqn9} due to the localized nature of measurement functions, then the above problem can be efficiently solved using probabilistic graphical modelling approach. The solution involves defining the FG corresponding to \eqref{eqn9}, and subsequently deriving expressions for BP messages exchanged over the FG, as detailed next.
\subsection{FG representation of bus/branch model}
In order to transform the bus/branch model into the FG, every state variable (bus voltage magnitudes and bus voltage angles) is represented as a variable node while every measurement is represented as a factor node. Links between variable nodes and factor nodes are defined according to the measurement functions, where each variable node is connected to the factor node if the variable is an argument of the measurement function. 

Measurements that directly measure state variables are referred to as \emph{direct} measurements, and those include $\mathbf z_{dir} \in \{\mathbf z_{V_i},\mathbf z_{\theta_i}\}$. Otherwise, we call \emph{indirect} measurements those that measure state variable indirectly such as $\mathbf{z}_{ind} \in \{\mathbf{z}_{P_{ij}}, \mathbf{z}_{Q_{ij}}, \mathbf{z}_{P_i}, \mathbf{z}_{Q_i}, \mathbf{z}_{I_{ij}}\}$. The corresponding factor nodes in the FG are denoted as $f_{dir} \in \{f_{V_i}, f_{\theta_i}\}$ and $f_{ind}\in \{{f}_{P_{ij}}, {f}_{Q_{ij}}, {f}_{P_i}, {f}_{Q_i}, {f}_{I_{ij}}\}$, respectively. Note that the relationship between indirect measurements and state variables is described using measurement functions.

Observe a part of the electric power grid that consists of two buses with direct $M_{dir}$ and indirect measurement device $M_{ind}$. 
\begin{figure}[H]
\centering
\includegraphics[width=50.562mm]{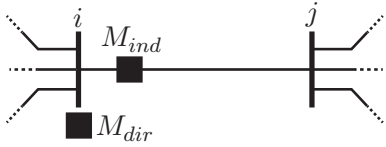}
\caption{Two buses part of an electric power grid}
\label{fig1}
\end{figure} \noindent
Input data for SE from these devices are Gaussian-type functions:
		\begin{equation}
        \begin{aligned}
		 M_{dir} &: \mathcal{N}(z_{dir}|x_i,\sigma_{dir}^2) \propto 
         \exp\Bigg\{\cfrac{(z_{dir}-x_{i})^2}{2\sigma_{dir}^2}\Bigg\} \\       
         M_{ind} &: \mathcal{N}(z_{ind}|\mathbf x,\sigma_{ind}^2) \propto 
         \exp\Bigg\{\cfrac{[z_{ind}-h_{ind}(\mathbf x)]^2}{2\sigma_{ind}^2}\Bigg\},         
        \end{aligned}
		\label{eqn12}
		\end{equation}
where measurement functions are defined as: $x_{i} \in \{V_i,\theta_i\}$ and $h_{ind}(\cdot) \in \{h_{P_{ij}}(\cdot), h_{Q_{ij}}(\cdot), h_{P_{i}}(\cdot), h_{Q_{i}}(\cdot), h_{I_{ij}}(\cdot)\}$, while variances $\sigma_{dir}^2$ and $\sigma_{ind}^2$ define errors of measurement devices. 
		
The BP algorithm on FGs proceeds by passing two types of messages along the edges of the FG: a variable node to a factor node and the factor node to a variable node messages.
Both variable and factor nodes in a FG process the incoming messages and calculate outgoing messages. As a general BP rule, an output message on any edge can be computed only upon reception of incoming messages from all other edges.

\subsection{Message from a variable node to a factor node}
\begin{figure}[!ht]
\centering
\includegraphics[width=6.8cm]{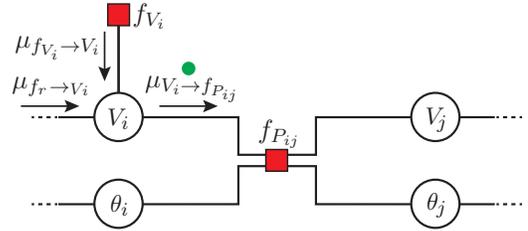}
\caption{Message from variable node to factor node}
\label{fig2}
\end{figure}

As an example, let us consider calculation of the message $\mu_{V_i \to f_{P{ij}}}$ as illustrated in Fig. \ref{fig2}. The direct measurement node $f_{V_i}$ will initialize and send the Gaussian message $\mu_{f_{V_i} \to V_i}\propto\mathcal{N}(z_{V_i}|V_i, \sigma_{V_i}^2)$ represented by a pair $(z_{V_i},\sigma_{V_i}^2)$ to the variable node $V_i$. Let us assume, for the time being, that the messages coming from the remaining edges of the graph are also Gaussian and represented by their corresponding mean-variance pairs. We generically denote such a message as $\mu_{f_r \to V_i}\propto\mathcal{N}(z_{f_r \to V_i}|V_i, \sigma_{f_r \to V_i}^2)$ and consider only a single such edge in Fig. \ref{fig2}. Note that this message carries the belief about the variable node $V_i$ as observed by its neighbouring factor node $f_r$. 

The message from a variable node to a factor node is equal to the product of all incoming factor node to variable node messages arriving at all the other incident edges. The resulting message represents the Gaussian function with mean $z_{V_i \to f_{P_{ij}}}$ and variance $\sigma_{V_i \to f_{P_{ij}}}^2$:  
		\begin{equation}
        \begin{aligned}
		\mu_{V_i \to f_{P_{ij}}} = \mu_{f_{V_i} \to V_i} \cdot \mu_{f_r \to V_i} 
		\propto \mathcal{N}(z_{V_i \to f_{P_{ij}}}|V_i,\sigma_{V_i \to f_{P_{ij}}}^2)		
        \end{aligned}
		\label{eqn14a} \nonumber
        \end{equation}
		\begin{equation}
        \begin{aligned}
		z_{V_i \to f_{P_{ij}}} &= \cfrac{z_{V_i}\sigma_{f_r \to V_i}^2+z_{f_r \to V_i}\sigma_{V_i}^2}{\sigma_{V_i}^2+\sigma_{f_r \to V_i}^2}\\
		\sigma_{V_i \to f_{P_{ij}}}^2&=\cfrac{\sigma_{V_i}^2\sigma_{f_r \to V_i}^2}{\sigma_{V_i}^2+\sigma_{f_r \to V_i}^2};		
        \end{aligned}
		\label{eqn14}
        \end{equation}        
 \noindent		
or in a more practical form:	
		\begin{equation}
        \begin{aligned}
		\cfrac{1}{\sigma_{V_i \to f_{P_{ij}}}^2}&=\cfrac{1}{\sigma_{V_i}^2} + \cfrac{1}{\sigma_{f_r \to V_i}^2}\\
		z_{V_i \to f_{P_{ij}}} &=\Bigg( \cfrac{z_{V_i}}{\sigma_{V_i}^2}+ \cfrac{z_{f_r \to V_i}}{\sigma_{f_r \to V_i}^2} \Bigg)\sigma_{V_i \to f_{P_{ij}}}^2. 
        \end{aligned}
		\label{eqn15}
		\end{equation}
		
To summarize, a general form of the message from a variable node $x$ to a factor node $f$ is: 
		\begin{equation}
        \begin{gathered}
        \mu_{x \to f} =\prod_{f_a \in \mathcal{F}\setminus f} \mu_{f_a \to x}
		\propto \mathcal{N}(z_{x \to f}|x,\sigma_{x \to f}^2) \\
		\cfrac{1}{\sigma_{x \to f}^2}= \sum_{f_a \in \mathcal{F}\setminus f}
		 \cfrac{1}{\sigma_{f_a \to x}^2};\;\;\;\;
        z_{x \to f} = \Bigg( \sum_{f_a \in \mathcal{F}\setminus f}
         \cfrac{z_{f_a \to x}}{\sigma_{f_a \to x}^2}\Bigg)\sigma_{x \to f}^2, 
        \end{gathered}
		\label{eqn16}
		\end{equation}	
where $\mathcal{F}\setminus f$ defines the set of factor nodes which are directly connected to the variable node $x$  excluding the factor node $f$. 
\subsection{Message from a factor node to a variable node}
\begin{figure}[!ht]
\centering
\includegraphics[width=6.5cm]{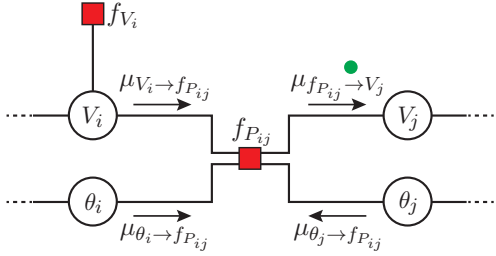}
\caption{Message from factor node to variable node}
\label{fig3}
\end{figure}

As an example, consider calculation of the message $\mu_{f_{P_{ij}} \to V_j}$, as shown in Fig. \ref{fig3}. The message can be computed only when all other incoming messages (variable to factor node messages) are known. As indicated, these messages are Gaussian functions, denoted as:  
		\begin{equation}
        \begin{gathered}
\mu_{V_i \to f_{P_{ij}}} \propto \mathcal{N}(z_{V_i \to f_{P_{ij}}}|V_i,\sigma_{V_i \to f_{P_{ij}}}^2)\\
\mu_{\theta_i \to f_{P_{ij}}} \propto \mathcal{N}(z_{\theta_i \to f_{P_{ij}}}|\theta_i,\sigma_{\theta_i \to f_{P_{ij}}}^2)\\
\mu_{\theta_j \to f_{P_{ij}}} \propto \mathcal{N}(z_{\theta_j \to f_{P_{ij}}}|\theta_j,\sigma_{\theta_j \to f_{P_{ij}}}^2).
        \end{gathered}
		\label{eqn17}
		\end{equation}	

The message from a factor node to a variable node is defined as a product of all incoming messages of the factor node multiplied by the Gaussian function associated to the factor node and marginalized over all of the variables associated with the incoming messages:
		\begin{equation}
        \begin{aligned}
		\mu_{f_{P_{ij}} \to V_j} &=\int\displaylimits_{V_i}\int\displaylimits_{\theta_i}\int\displaylimits_{\theta_j} 
		\mathcal{N}(z_{P_{ij}}|V_i,\theta_i,V_j,\theta_j,\sigma_{P_{ij}}^2) \\
		& \cdot \mu_{V_i \to f_{P_{ij}}}\cdot
		 \mu_{\theta_i \to f_{P_{ij}}}
		\cdot \mu_{\theta_j \to f_{P_{ij}}} \mathrm{d}V_i \mathrm{d}\theta_i \mathrm{d}\theta_j \\
		&\propto \mathcal{N}(z_{f_{P_{ij}} \to V_j}|V_j,\sigma_{f_{P_{ij}} \to V_j}^2).
        \end{aligned}
		\label{eqn18}
		\end{equation}
The result is the Gaussian function with mean $z_{f_{P_{ij}} \to V_j}$ and variance $\sigma_{f_{P_{ij}} \to V_j}^2$:
		\begin{equation}
		\begingroup\makeatletter\def\f@size{9}\check@mathfonts
		  \begin{gathered}
    z_{f_{P_{ij}} \to V_j}=\cfrac{z_{P_{ij}}-g_{ij}z_{V_i \to f_{P_{ij}}}+b_{ij}z_{\theta_i \to f_{P_{ij}}}-b_{ij}z_{\theta_j \to f_{P_{ij}}}}{-g_{ij}}\\
		\sigma_{f_{P_{ij}} \to V_j}^2 = \cfrac{\sigma_{P_{ij}}^2+g_{ij}^2\sigma_{V_i \to f_{P_{ij}}}^2+b_{ij}^2\sigma_{{\theta_i \to f_{P_{ij}}}}^2+b_{ij}^2\sigma_{{\theta_j \to f_{P_{ij}}}}^2}{g_{ij}^2}.
		\label{eqn19}
		\end{gathered}
		\endgroup
		\end{equation}
The parameters $g_{ij}$ and $b_{ij}$ are defined according to the indirect measurement function $ h_{P_{ij}}(\cdot)$. Given that the input initialization messages to the BP algorithm are Gaussians, and that both variable and factor node processing preserve Gaussian form of messages, the presented BP for extended DC model is an instance of Gaussian BP \cite{ping}.

To summarize, a general form of the message from a factor node $f$ to a variable node $x$ is:
		\begin{equation}
        \begin{gathered}
        \mu_{f \to x}= 
		\int\displaylimits_{x_1}\dots\int\displaylimits_{x_p}
		\mathcal{N}(z_{f}|x,x_1\dots x_p,\sigma_{f}^2)\\
		\prod_{x_b \in \mathcal{X}\setminus x} \mu_{x_b \to f} \cdot \mathrm{d}x_b \propto \mathcal{N}(z_{f \to x}|x,\sigma_{f \to x}^2)\\
		z_{f \to x}=\cfrac{z_f-C_1z_{x_1 \to f}-\dots -C_{p}z_{x_p \to f}-K }{C}\\
		\sigma_{f \to x}^2 = \cfrac{\sigma_f^2+C_1^2\sigma_{x_1 \to f}^2+\dots +C_{p}^2\sigma_{x_p \to f}^2}{C^2},
        \end{gathered}
		\label{eqn21}
		\end{equation}
where $\mathcal{X}\setminus x=\{x_1,\dots, x_p\}$ are the set of variable nodes incident to the factor node $f$, excluding the variable node $x$. The coefficients $C,C_1,...,C_p,K$ are defined according to the measurement function associated with the factor node with mean $z_f$ and variance $\sigma_f^2$:		
		\begin{equation}
        \begin{gathered}
		f(x,x_1\dots x_p)=Cx+C_1x_1\dots +C_px_p+K.
		\end{gathered}
		\label{eqn22a}
		\end{equation}
		
\subsection{Marginals}
The marginals of each state variable are obtained as the product of all incoming messages into the variable node. Thus the resulting marginals are Gaussians with mean and variance calculated as in \eqref{eqn16}, except that the product includes all terms $f_a \in \mathcal{F}$.

\subsection{Convergence}
It is well known that loopy BP does not always converge to correct marginals. Our numerical studies show that the convergence of BP algorithm strongly depends on measurement data, where specific inputs may lead to oscillatory behaviour of messages. After extensive numerical analysis, the following heuristic solution is adopted to improve the convergence of the BP algorithm \cite{pretti}. We modify updates of factor to variable node messages $\mu_{f \to x}^k$ in the current (k-th) iteration as follows:
		\begin{equation}
        \begin{gathered}
		  \mu_{f \to x}^k=[1-\delta(p)]\cdot\mu_{f \to x}^k + \delta(p)\cdot \alpha \cdot [\mu_{f \to x}^{k-1}+\mu_{f \to x}^k], 
		\end{gathered}
		\label{eqn22}
		\end{equation}
where $\delta(p) \in \{0,1\}$ is Bernoulli random variable with parameter $p$, independently sampled for each message $\mu_{f \to x}^k$, and $\alpha$ is weighting coefficient.  For the values of $p\in [0.4,0.6]$ and $\alpha=0.5$, our numerical studies show that the BP algorithm always converges successfully to the correct solution.

\section{Numerical Example}
Fig. 4 shows the IEEE 14 bus test case. The available measurement devices are: active $M_{P_{ij}}$ and reactive $M_{Q_{ij}}$ power flow, injection active $M_{P_{i}}$ and reactive $M_{Q_{i}}$ power, voltage magnitude $M_{V_{i}}$ and angle $M_{\theta_{i}}$. Measurement devices have readings $z_{P_{ij}}$, $z_{Q_{ij}}$, $z_{P_{i}}$, $z_{Q_{i}}$, $z_{V_{i}}$ and $z_{\theta_{i}}$ with variances $\sigma_{P_{ij}}^2$, $\sigma_{Q_{ij}}^2$, $\sigma_{P_{i}}^2$, $\sigma_{Q_{i}}^2$, $\sigma_{V_{i}}^2$ and $\sigma_{\theta_{i}}^2$, respectively. 

\begin{figure}[!ht]
\centering
\includegraphics[width=6.7cm]{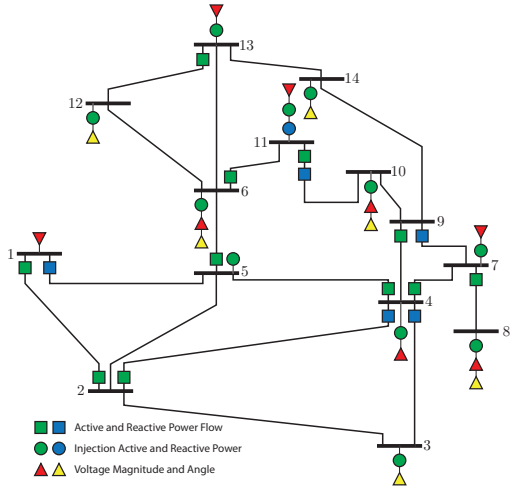}
\caption{The IEEE 14 bus test case}
\label{fig4}
\end{figure}

The set of measurements is generated using extended DC power flow analysis, additionally corrupted by Gaussian white noise. Using Monte Carlo approach, we generate 1000 random sets of
measurement values for different values of measurement variances from the set $\sigma^2=$ $\{\sigma_1^2$, $\sigma_2^2$, $\sigma_3^2 \}$ $=\{0.01^2$, $0.001^2$, $0.0001^2 \} \; [\mbox{p.u.}]$, and fed them to the proposed BP-based SE algorithm in order to obtain the average performance results. The parameter which is used to evaluate convergence behaviour is root mean square error (RMSE) defined as $\mathrm{RMSE} = { \frac{1}{2n} ||\mathbf {\hat x}- \mathbf{x}||_2 }$,
where $\mathbf {\hat x}$ is weighted least-squares solution, while $\mathbf{x}$ represents the solution of the BP algorithm.

The iterative BP algorithm is applied as follows:
\begin{enumerate}
	\item all factor nodes associated to direct measurements send messages to corresponding variable nodes;
	\item all variable nodes send messages along incidence edges (except to an edge towards a factor node associated to direct measurement) the form of the messages is:
	\begin{enumerate}
		\item messages are equal to the message from step 1, if variable nodes have direct measurements; 
		\item messages take the form of the "flat start" given by distribution with means $V_i=1$ or $\theta_i=0$ and variances $\sigma_{V_i}^2 \to \infty$ or $\sigma_{\theta_i}^2 \to \infty$, if variable nodes do not have direct measurements;  
	\end{enumerate}
	\item all factor nodes compute messages to incident variable nodes according to \eqref{eqn21};
	\item all variable variable nodes compute messages to incident factor nodes according to \eqref{eqn16};
	\item all variable nodes compute corresponding marginal distributions;
	\item repeat steps 3, 4, 5 until BP converges. 
\end{enumerate}

\begin{figure}[!ht]
\centering
\includegraphics[width=6cm]{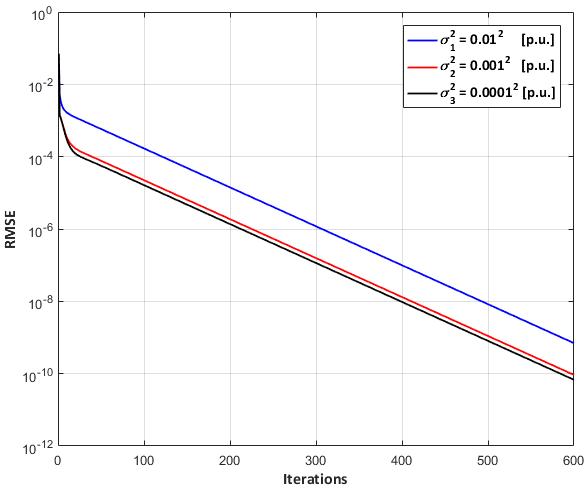}
\caption{The IEEE 14 convergence performance for variances $\sigma^2$}
\label{fig5}
\end{figure}

Fig. \ref{fig5} shows the convergence of the proposed BP towards the weighted least-squares solution for IEEE 14 bus case presented in Fig. \ref{fig4}. We note that the BP solution converges to the weighted least-squares solution for a range of noise variances within several hundreds of iterations. We obtained similar curves for IEEE 30 bus test case. In fact, comparing our results with the results presented in \cite{raffi}, the BP algorithm has comparable number of iterations with the best performing distributed algorithms analysed therein. Note that \cite{raffi} analyses multi-area SE problem with four areas defined on the IEEE 14 network, while we are dealing here with fully distributed case which is expected to have slower convergence.

\section{Conclusion}
In this paper, we provided FG representation and applied the BP algorithm to efficiently evaluate the SE solution for the extended DC model. The generic format of BP messages is presented that will be preserved (albeit somewhat extended) for the non-linear AC model; as demonstrated in our follow up work. 


\section*{Acknowledgment}
This project has received funding from the European Union’s Seventh Framework Programme for research, technological development and demonstration under grant agreement no. 607774.

\bibliographystyle{IEEEtran}
\bibliography{mybib}

\begin{thebibliography}{10}
\providecommand{\url}[1]{#1}
\csname url@samestyle\endcsname
\providecommand{\newblock}{\relax}
\providecommand{\bibinfo}[2]{#2}
\providecommand{\BIBentrySTDinterwordspacing}{\spaceskip=0pt\relax}
\providecommand{\BIBentryALTinterwordstretchfactor}{4}
\providecommand{\BIBentryALTinterwordspacing}{\spaceskip=\fontdimen2\font plus
\BIBentryALTinterwordstretchfactor\fontdimen3\font minus
  \fontdimen4\font\relax}
\providecommand{\BIBforeignlanguage}[2]{{%
\expandafter\ifx\csname l@#1\endcsname\relax
\typeout{** WARNING: IEEEtran.bst: No hyphenation pattern has been}%
\typeout{** loaded for the language `#1'. Using the pattern for}%
\typeout{** the default language instead.}%
\else
\language=\csname l@#1\endcsname
\fi
#2}}
\providecommand{\BIBdecl}{\relax}
\BIBdecl

\bibitem{abur}
A.~Abur and A.~Exp{\'o}sito, \emph{Power System State Estimation: Theory and
  Implementation}, ser. Power Engineering.\hskip 1em plus 0.5em minus
  0.4em\relax Taylor \& Francis, 2004.

\bibitem{monticelliPaper}
A.~Monticelli, ``Electric power system state estimation,'' \emph{Proc. of the
  IEEE}, vol.~88, no.~2, pp. 262--282, Feb 2000.

\bibitem{pearl}
J.~Pearl, \emph{Probabilistic Reasoning in Intelligent Systems: Networks of
  Plausible Inference}.\hskip 1em plus 0.5em minus 0.4em\relax San Francisco,
  CA, USA: Morgan Kaufmann Publishers Inc., 1988.

\bibitem{bishop}
C.~M. Bishop, \emph{Pattern Recognition and Machine Learning}.\hskip 1em plus
  0.5em minus 0.4em\relax Springer, 2006.

\bibitem{kavcic}
Y.~Hu, A.~Kuh, A.~Kavcic, and D.~Nakafuji, ``Real-time state estimation on
  micro-grids,'' in \emph{IJCNN, The 2011 International Joint Conference on},
  July 2011, pp. 1378--1385.

\bibitem{ilic}
Y.~Weng, R.~Negi, and M.~Ilic, ``Graphical model for state estimation in
  electric power systems,'' in \emph{SmartGridComm, 2013 IEEE International
  Conference on}, Oct 2013, pp. 103--108.

\bibitem{koster}
A.~Koster and S.~Lemkens, ``Network optimization: Designing ac power grids
  using integer linear programming,'' \emph{LNCS, Springer}, pp. 478--483,
  2011.

\bibitem{barber}
D.~Barber, \emph{{Bayesian Reasoning and Machine Learning}}.\hskip 1em plus
  0.5em minus 0.4em\relax {Cambridge University Press}, 2012.

\bibitem{ping}
H.~A. Loeliger, J.~Dauwels, J.~Hu, S.~Korl, L.~Ping, and F.~R. Kschischang,
  ``The factor graph approach to model-based signal processing,'' \emph{Proc.
  of the IEEE}, vol.~95, no.~6, pp. 1295--1322, 2007.

\bibitem{pretti}
M.~Pretti, ``A message-passing algorithm with damping,'' \emph{Journal of
  Statistical Mechanics: Theory and Experiment}, vol. 2005, no.~11, 2005.

\bibitem{raffi}
\BIBentryALTinterwordspacing
R.~Sevlian and U.~Ponsukcharoen, ``Distributed power system state estimation,''
  Technical Report, June 2012. [Online]. Available:
  \url{http://web.stanford.edu/~rsevlian/raffi\_sevlian\_papers/CEE272R.pdf}
\BIBentrySTDinterwordspacing

\end{thebibliography}
\end{document}